\documentclass[prc,aps]{revtex4}
\draft
\usepackage{graphicx}
\usepackage{color}
\usepackage{mathtools}
\begin{document}

\title{Exploring the relationship between nuclear matter and finite nuclei with chiral two- and three-nucleon forces}       
\author{            
Francesca Sammarruca  and Randy Millerson  }                                                           
\affiliation{ Physics Department, University of Idaho, Moscow, ID 83844-0903, U.S.A. 
}
\date{\today} 
\begin{abstract}
We address the connection between the saturating behavior of infinite nuclear matter and the description of finite nuclei based on state-of-the-art chiral two- and three-nucleon forces. We observe that chiral two- and three-nucleon interactions (at N$^2$LO and at N$^3$LO) which have been found to predict realistic binding energies and radii for a wide range of finite nuclei (from p-shell nuclei up to nickel isotopes) are unable to saturate infinite nuclear matter. On the other hand, it has been shown that, when the fits of the $c_D$ and $c_E$ couplings of the chiral three-nucleon interactions include the constraint of nuclear matter saturation in addition to, as is typically the case, the triton binding energy, medium-mass nuclei are underbound and their radii are sytematically too large. We discuss this apparent inconsistency and perform test calculations for various scenarios to shed light on the issue.

\end{abstract}
\maketitle 
        
\section{Introduction} 
\label{Intro} 

Understanding the interaction of hadrons in nuclei is one of the most fundamental problems in             
nuclear physics. Our present knowledge of the nuclear force in vacuum is still incomplete, although 
decades of efforts have been devoted to this problem. The study of nuclear forces in many-body systems is even more challenging because additional aspects are involved beyond those which can be constrained by free-space nucleon-nucleon ($NN$) scattering. Predictive power with respect to the properties of nuclei     
is the true test for a successful microscopic theory.      

 Traditionally, the system known as infinite nuclear matter has been considered to be the ``test bench" for nuclear many-body theories. Nuclear matter is defined as an infinite system of nucleons interacting {\it via} nuclear forces in the absence of electromagnetic interactions. Nuclear matter is characterized by its equation of state (EoS), namely the energy per particle as a function of density (and other quantities as appropriate, such as isospin asymmetry). The idealized nature of this system, which implies translational invariance, simplifies theoretical calculations. Furthermore, {\it via} the ``local density approximation" (LDA), one can utilize the EoS directly in calculations of actual nuclei. (We recall that LDA amounts to the assumption that the properties at a point in a nucleus with density $\rho$ are the same as they are in infinite nuclear matter of the same density.)

 As nuclear matter is the result of an extrapolation from finite nuclei, the saturation properties of symmetric nuclear matter (SNM), that is, the minimum of the EoS at the appropriate equilibrium density, should be naturally related to the energy and density distributions of nucleons in nuclei. Typically, the extrapolation from  finite nuclei to SNM has been done with phenomenological density-dependent forces, such as Skyrme or Gogny forces~\cite{Gogny,Skyrme,Skyrme-Gogny}. Applying such forces, it has been observed that a good phenomenological description of nuclei extrapolates to a saturation density in nuclear matter of about 0.16 fm$^{-3}$ and energy per particle of approximately -16 MeV. 

Alternatively, one may ask the question:  what do {\it microscopic} approaches predict with regard to such relationship?
It is the main purpose of this paper to address this question and examine the connection between the EoS of SNM and the description of finite nuclei for microscopic approaches. We will consider {\it ab initio} calculations which employ chiral two- and three-nucleon forces. Also, we will perform approximate calculations of finite nuclei based on a mass formula applying an EoS microscopically obtained from chiral interactions.This will give us the opportunity to explore the aforementioned connection from another angle. We also note the work of Ref.~\cite{Atk+}, where the relationship between nuclear matter and finite nuclei is investigated using a dispersive optical model and self-consistent Green's functions.

Over the past several years, chiral effective field theory (EFT) has evolved into the most favorable approach  for constructing nuclear interactions. It provides for a systematic way to construct 
nuclear many-body forces, which emerge on an equal footing~\cite{Wei92} with two-body forces, and 
to assess theoretical uncertainties through an expansion controlled by an organizational scheme known as
``power counting"~\cite{Wei90}. Furthermore, chiral EFT maintains consistency with the underlying fundamental theory of strong interactions, quantum chromodynamics (QCD), through symmetries and
symmetry breaking patterns.       
     
This paper is organized as follows. First, we review the main aspects of our calculations of SNM, which include chiral three-nucleon forces (3NFs) up to next-to-next-to-next-to-leading order (N$^3$LO), see Sec.~\ref{calc}. In Sec.~\ref{res}, we present and discuss results for the EoS and turn to the literature for {\it ab initio} predictions of finite nuclei with similar few-nucleon forces. In Sec.~\ref{nuc}, we explore further the relation between the saturation of SNM and the properties of some selected nuclei.
Our observations and conclusions are summarized in Sec.~\ref{Concl}. 

\section{Description of the calculations} 
\label{calc}

 We perform microscopic calculations of nuclear matter using the nonperturbative particle-particle ladder approximation, which generates the leading-order contributions in the traditional hole-line expansion. We compute the single-particle spectrum for the intermediate state also in the particle-particle ladder approximation, keeping only the real part. For a recent review of our nuclear matter calculations, see Ref.~\cite{SM_front}.

The input two-nucleon forces (2NFs) and 3NFs are described next.
 
\subsection{The two-nucleon forces}  
\label{II} 

The $NN$ potentials employed in this work are part of a set which spans five orders in the chiral EFT expansion, from leading order (LO) to fifth order (N$^4$LO)~\cite{EMN17}. For the construction of these potentials, the same power counting scheme and regularization procedures are applied through all orders, making this set of interactions more consistent than previous ones.  Another novel and important aspect in the construction of these new potentials is the fact that the long-range part of the interaction is fixed by the $\pi N$ low-energy constants (LECs) as determined in the recent and very accurate analysis of Ref.~\cite{Hofe+}. In fact, for all practical purposes, errors in the $\pi N$ LECs are no longer an issue with regard to uncertainty quantification. Furthemore, at the fifth (and highest) order, the $NN$ data below pion production threshold are reproduced with excellent precision ($\chi ^2$/datum = 1.15).

Iteration of the potential in the Lippmann-Schwinger equation requires cutting off high-momentum components, consistent with the fact that chiral perturbation theory amounts to building a low-momentum expansion. This is accomplished through the application of a regulator function for which the non-local form is chosen:
\begin{equation}
f(p',p) = \exp[-(p'/\Lambda)^{2n} - (p/\Lambda)^{2n}] \,,
\label{reg}
\end{equation}
where $p' \equiv |{\vec p}\,'|$ and $p \equiv |\vec p \, |$ denote the final and initial nucleon momenta in the two-nucleon center-of-mass system, respectively. In the present applications, we will 
consider values for the  cutoff parameter $\Lambda \leq 500$ MeV. 
The potentials are relatively soft as confirmed by the 
Weinberg eigenvalue analysis of Ref.~\cite{Hop17} and in the context of the perturbative calculations of infinite matter of  Ref.~\cite{DHS19}.

\subsection{The three-nucleon forces} 
\label{III}

Three-nucleon forces first appear at the third order of the chiral expansion (N$^2$LO). At this order, the 3NF consists of three contributions~\cite{Epe02}: the long-range two-pion-exchange (2PE) graph, the medium-range one-pion exchange (1PE) diagram, and a short-range contact term. 

For nuclear matter calculations, these 3NFs can be expressed as density-dependent effective two-nucleon interactions as derived in Refs.~\cite{holt09,holt10}. They are represented in  terms of the well-known non-relativistic two-body nuclear force operators and, therefore, can be conveniently incorporated in the usual $NN$ partial wave formalism and the particle-particle ladder approximation for computing the EoS.       
         
The effective density-dependent two-nucleon interactions at N$^2$LO consist of six one-loop topologies. Three of them are generated from the 2PE graph of the chiral 3NF and depend on the LECs $c_{1,3,4}$, which are already present in the 2PE part of the $NN$ interaction. Two one-loop diagrams are generated from the 1PE diagram, and depend on the low-energy constant $c_D$. Finally, there is the one-loop diagram that involves the 3NF contact diagram, with LEC $c_E$. 

The complete 3NF beyond N$^2$LO is very complex and was neglected in nuclear structure studies of the past.
 However, in recent years, the 3NF at N$^3$LO has been derived~\cite{Ber08} and applied in some nuclear many-body systems~\cite{Tew13,Dri16,DHS19,Heb15a}. 
The contributions to the subleading chiral 3NF include: the 2PE topology, which is the longest-range component of the subleading 3NF, the two-pion-one-pion exchange topology, and the ring topology, generated by a circulating pion which is absorbed and reemitted from each of the three nucleons. 

Direct inclusion of the subleading chiral 3NF is very challenging for many-body calculations. However, similar to the leading 3NF, the contributions of the 3NF at N$^3$LO can be conveniently expressed in the form of density-dependent effective two-nucleon interactions, as derived in Refs.~\cite{Kais18,Kais19}. Here, we retain all the long-range components~\cite{Kais19}. 

The in-medium $NN$ potentials corresponding to the short-range subleading 3NFs have been calculated in Ref.~\cite{Kais18}, and include the two-pion-exchance-contact topology and the relativistic corrections, proportional to $1/M$, where $M$ is the nucleon mass. Both have been shown to be negligible~\cite{Tew13, Heb15a} and are therefore left out in this study.

\section{Results for the Equation of State}
\label{res}

\begin{table*}
\caption{Values of the LECs used for the 3NFs applied in this work.}
\label{lecs}
\centering
\begin{tabular*}{\textwidth}{@{\extracolsep{\fill}}ccccc}
\hline
\hline
Source & chiral order & $\Lambda$ (MeV) & $c_{D}$  & $c_{E}$  \\
\hline
\hline
Ref.~\cite{DHS19}  & N$^2$LO & 450 & {\bf (a)}  2.25 & 0.07  \\ 
    &     &  & {\bf (b)} 2.50 & 0.1   \\       
 &      &  &  {\bf (c)} 2.75  & 0.13   \\       
      &       & 500 & {\bf (a)}  -1.75  &   -0.64  \\ 
    &     &  & {\bf (b})  -1.50  &  -0.61   \\       
 &      &  & {\bf (c)}  -1.25  &  -0.59  \\      
   &  N$^3$LO   & 450  &0.0  &  -1.32  \\       
 &    &   & 0.25  &   -1.28   \\       
&    &   & 0.50  &  -1.25  \\    
  &     & 500  & -3.00  &   -2.22   \\       
 &    &   & -2.75  &  -2.19  \\       
&    &   & -2.50,  &  -2.15  \\    
\hline
Ref.~\cite{Huether+2020}  & N$^2$LO & 450 & 10.0  &   0.909   \\ 
      &       & 500 & 5.0   &   -0.159   \\ 
   &  N$^3$LO   & 450  & 9.0  &   -0.152   \\        
  &     & 500  & 4.0  &   -1.492   \\       
\hline
\hline
\end{tabular*}
\end{table*}

With the tools described above, we proceed to calculate the EoS of SNM.

\begin{figure*}[!t] 
\centering
\hspace*{-3cm}
\includegraphics[width=8.7cm]{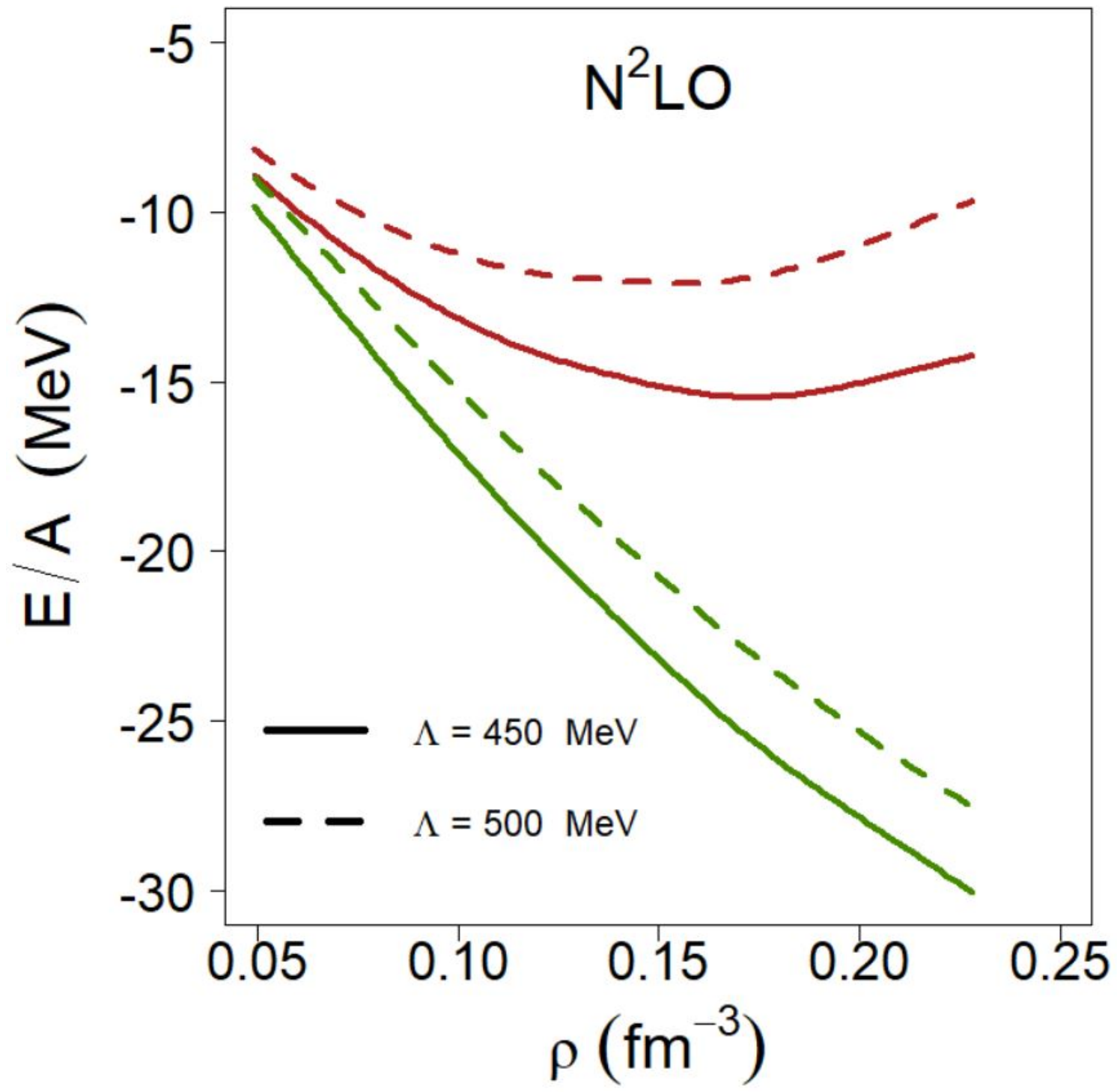}\hspace{0.01in} 
 \includegraphics[width=8.7cm]{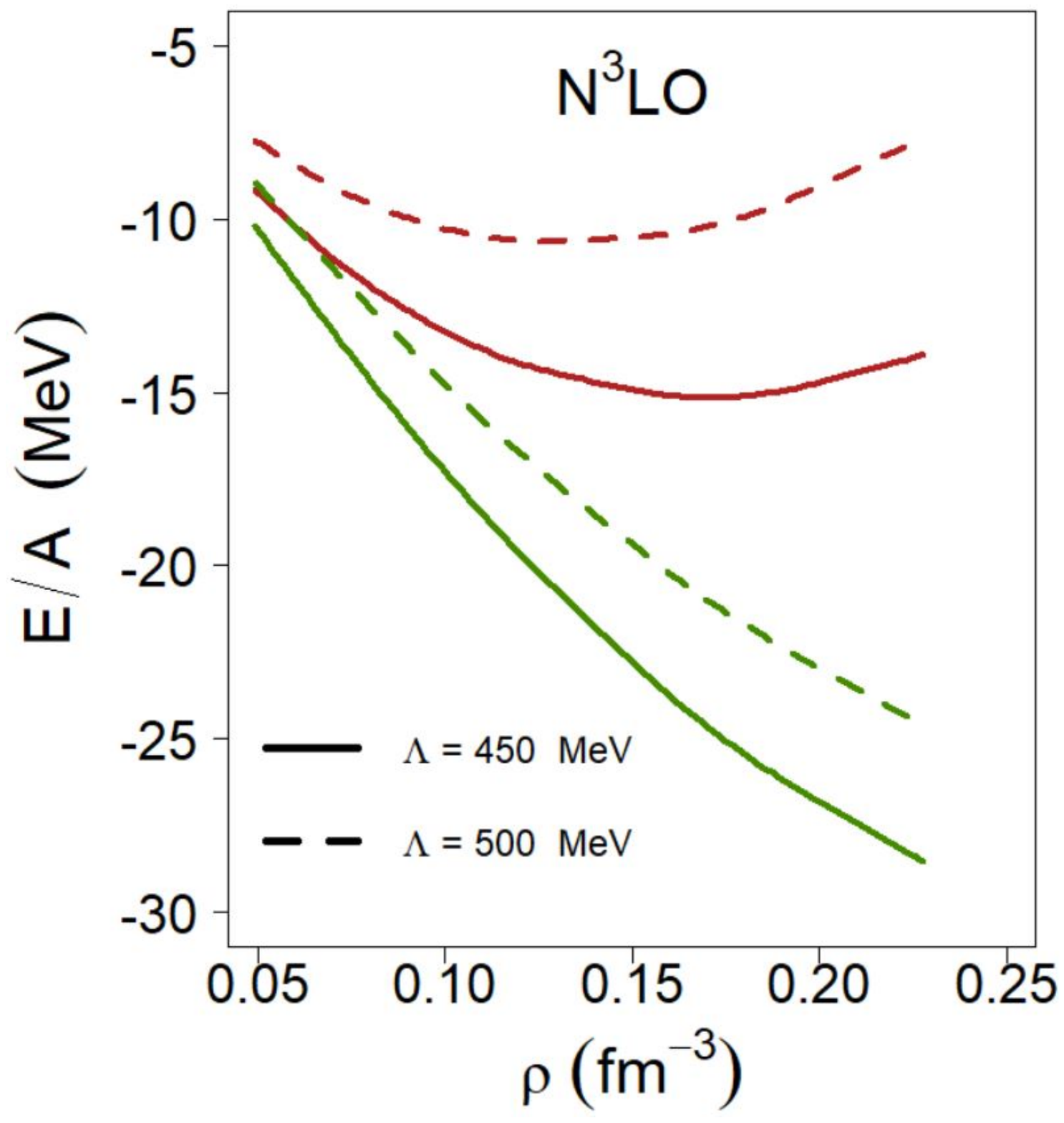}\hspace{0.01in} 
\vspace*{-2.5cm}
 \caption{(Color online) Energy per particle in SNM as a function of density. Left side: All calculations include the 2NF and the 3NF at N$^2$LO. The upper (red) curves apply the $c_D$, $c_E$ from Ref.~\cite{DHS19} (case (b)), whereas  the lower (green) curves use the $c_D$, $c_E$ from Ref.~\cite{Huether+2020}. Right:  same as left side, but with the 2NF and the 3NF at N$^3$LO. Concerning the values for the $c_D, c_E$ LECs applied in the 3NF, see Table~\ref{lecs}. } 
 \label{n2lo_n3lo}
\end{figure*}

We begin with discussing Fig.~\ref{n2lo_n3lo}, where we show the EoS at N$^2$LO including the 3NF at N$^2$LO (left side), and the EoS at N$^3$LO including the 3NF at N$^2$LO plus the long-range 3NF at  N$^3$LO (right side). In both the left and the right frames, 
for the two upper (red) curves,  the low-energy constants (LECs) $c_E$, $c_D$ of Ref.~\cite{DHS19} are used, see Table~\ref{lecs}. 
Those values were obtained as follows. First, the authors performed fits to the $^3$H binding energy, which leads to a relation between $c_D$ and $c_E$. For both values of the cutoff, they found $c_E$ couplings of natural size within a wide range of $c_D$. Following these trajectories, they then identified acceptable ($c_D,c_E$) pairs from calculations of nuclear matter, where they obtained the saturation point as a function of $c_D$. The final choices are those most consistent with both the empirical saturation energy and density calculated at fourth order of many-body perturbation theory (MBPT). For more details, see Supplemental Material cited in Ref.~\cite{DHS19}. The same procedures and considerations were applied both at N$^2$LO and N$^3$LO.

We find good agreement between our predictions with $\Lambda$=450 MeV, which is our standard choice, and the corresponding EoS predictions of Ref.~\cite{DHS19}, especially considering that we use a different many-body method (non-perturbative Brueckner-Hartree-Fock (BHF) instead of MBPT) and we calculate the 3NF contribution by way of a density-dependent 2NF, see Sec.~\ref{III}. At saturation, we estimate the differences to be in the order of 2 and 3\% at N$^3$LO and N$^2$LO, respectively. The differences for $\Lambda$=500 MeV are considerably larger, which could be attributed, in part, to a less desirable perturbative behavior for the larger cutoff.

Moving to the green curves of Fig.~\ref{n2lo_n3lo}, the scenario is dramatically different. Those EoS were obtained with the 3NF LECs from Ref.~\cite{Huether+2020}, which were fitted using $^3$H and $^{16}$O ground-state energies. In that way, an excellent reproduction of both experimental energies and radii from p-shell nuclei up to the nickel isotopes was achieved. However, the EoS we obtain with those couplings are clearly overly attractive and show no sign of saturation.

On the other hand, the interactions constructed in Ref.~\cite{DHS19}, corresponding to the red curves in both frames of Fig.~\ref{n2lo_n3lo},  lead to underbound ground-state energies for finite nuclei and radii which are systematically too large~\cite{Hoppe19}.

\begin{figure}[!t] 
\centering
\vspace*{-1inch}\includegraphics[width=8.7cm]{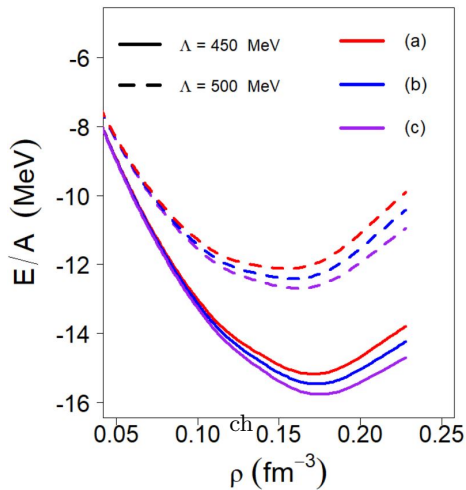}\hspace{0.01in} 
\vspace*{-2.5cm}
\caption{(Color online) Energy per particle in SNM as a function of density at N$^2$LO for $\Lambda$=450 MeV and $\Lambda$=500 MeV. See Table~\ref{lecs} for the various values of the $c_D$, $ c_E$ applied in the 3NF.} 
 \label{dr_all}
\end{figure}

In Fig.~\ref{dr_all}, we present in more detail the calculations of Ref.~\cite{DHS19} at N$^2$LO.
We compare the EoS at N$^2$LO using, for each cutoff, all three different $(c_D,c_E)$ combinations found in Ref.~\cite{DHS19} (cf.~Table~\ref{lecs}).
Case (c) of $\Lambda$=450 MeV displays the best saturating behavior. 

The above observations constitute the problem we wish to address in this paper.

Constraints from the few-nucleon system and a relatively light nucleus such as $^{16}$O~\cite{Huether+2020} produce chiral interactions which are excessively attractive when applied in nuclear matter.  On the other hand, simultaneous constraints from both the few-nucleon system and the SNM saturation point produce underbound medium-mass nuclei and an EoS that is mostly on the repulsive side. If the triton binding energy constraint is removed, changes in the 3NF couplings (especially the short-range one) allowed an improved description of medium-mass nuclei while substantially overbinding the triton~\cite{Hoppe19}.

It appears that the 3NF behaves very differently in infinite matter as compared to finite nuclei. Note that this puzzling scenario is independent of whether the 3NF is applied in the form of a density-dependent effective two-body potentials as we do, or in MBPT as done by Drischler {\it et al.}~\cite{DHS19}.

\begin{figure}[!t] 
\centering
\vspace*{-1inch}\includegraphics[width=8.7cm]{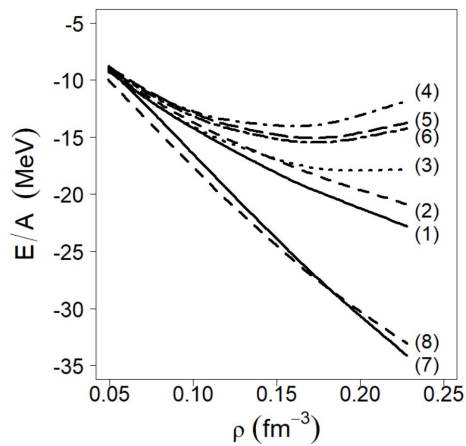}\hspace{0.01in} 
\vspace*{-3.0cm}
 \caption{ Energy per particle in SNM as a function of density demonstrating the individual contributions from the 3NF.  All calculations are performed at N$^2$LO with cutoff equal to 450 MeV. Curve (1): no 3NF included (i.~e. 2NF only); curves (2), (3), (4) include 3NF contributions proportional to $c_1$,   $c_3$, and $c_4$, respectively, added up successively. Curve (5) includes, in addition, the contributions proportional to $c_D$. All contributions are contained in curve (6) (including the $c_E$ contribution). For curves (5) and (6), the couplings (b) from Ref.~\cite{DHS19} are used (cf. Table~\ref{lecs}),  whereas for curves (7) and (8)
 those from Ref.~\cite{Huether+2020} are adopted.} 
 \label{3nf}
\end{figure}

The large sensitivity of the $c_D$, $c_E$ LECs to the systems/properties used to constrain their values, apparent from Table~\ref{lecs}, is just another way to state the same puzzle. In an effort to shed more  light on this interesting question, we calculate the EoS for the individual 3NF contributions, shown in Fig.~\ref{3nf}. Since the problem we are discussing is apparently independent of whether the calculations are conducted at third or fourth order of the chiral expansion or the value of the cutoff, we choose N$^2$LO with  $\Lambda$=450 MeV as our demonstration case.

In Fig.~\ref{3nf}, we start from a baseline EoS with only the 2NF, curve (1), and then add 3NF contributions one by one. Curves (2) to (4) are obtained by including the contributions proportional to $c_1$ (curve (2)), $c_1$ and $c_3$ (curve (3)), $c_1$, $c_3$, and $c_4$ (curve (4)). The curve labeled (5) includes, in addition, the contributions proportional to $c_D$, while curve (6) contains all 3NF contributions at N$^2$LO (i.~e., also the $c_E$ contribution). Contributions are added up succesively.  For (5) and (6), the values for $(c_D,c_E)$ are  those of Ref.~\cite{DHS19} (case (b)). The curves labeled (7) and (8) are obtained with the $(c_D,c_E)$ 3NF couplings used in Ref.~\cite{Huether+2020} (solid green curve in Fig.~\ref{n2lo_n3lo}, left side). 

From Fig.~\ref{3nf}, we see that the term proportional to $c_1$ is small and repulsive, and that the $c_3$-contribution  provides a hint of saturation. The $c_4$-term is instrumental for saturation, while both $c_D$ and $c_E$ add attraction. 

The figure also confirms that the large value of $c_D$ applied for curve (7) is responsible for the non-saturating behavior.
As both sets of 3NF couplings applied in curves (5) and (6) {\it vs.} (7) and (8) are consistent with the triton binding energy, one may conclude that $c_D$ has a much larger impact in nuclear matter than in the three-nucleon system, confirming the observation in Ref.~\cite{Hoppe19}.

\begin{figure}[!t] 
\centering
\vspace*{-1inch}\includegraphics[width=8.7cm]{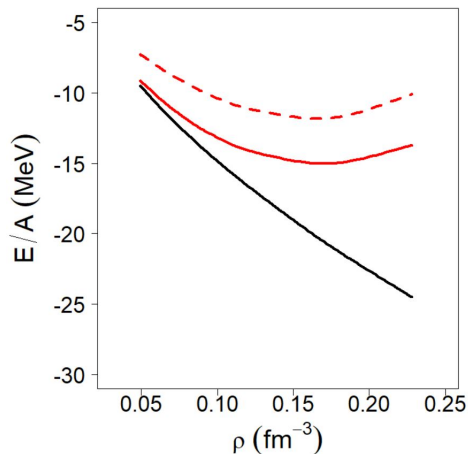}\hspace{0.01in} 
\vspace*{-3.0cm}
 \caption{ Energy per particle in SNM as a function of density demonstrating the individual contributions from the 3NF at N$^2$LO and N$^3$LO.  All calculations are performed with cutoff equal to 450 MeV. Solid black: only the 2NF at N$^3$LO is included; solid red: total 3NF at N$^3$LO is included; dashed red: only the N$^2$LO part of the 3NF at N$^{3}$LO is included.} 
 \label{n2n3}
\end{figure}

Before closing this section, we wish to identify the contribution from the N$^3$LO portion of the 3NF at N$^3$LO, see Fig.~\ref{n2n3}. In this figure, the solid black curve  is the result when only the 2NF at N$^3$LO (with cutoff equal to 450 MeV) is applied.  The solid red curve includes the entire N$^3$LO 3NF, while the red dashed curve includes only the N$^2$LO part of the N$^3$LO 3NF. Thus, the difference between the red dashed and the red solid curves represents the N$^3$LO portion of the 3NF,  which is moderately attractive. Note that the N$^3$LO contribution to the 3NF is parameter-free~\cite{Ber08,Kais19} and, therefore, this is a general result. It is consistent with the findings of Ref.~\cite{Heb15a}.

\section{Elucidating the connection between nuclear matter and finite nuclei}
\label{nuc}

In the previous section, we have confirmed that, when $c_D, c_E$ are determined through the ground-state energy of a nucleus such as $^{16}$O, the resulting values produce way too much attraction in saturated SNM, a system with density approximately equal to $0.16$ fm$^{-3}$. Vice versa, values constrained by the saturation properties of SNM underbind $^{16}$O. This ``mismatch"~\cite{Huether+2020}, while not understood, may be seen as an indication that the chiral 3NF operates differently for systems with different densities or density distributions.

An intuitive picture, established in nuclear physics since decades, describes a nucleus in terms of a mass formula, whose extrapolation to an infinite electrically neutral system is known as nuclear matter. Although simple, this model should not be fundamentally wrong, especially for bulk properties such as energies and r.m.s.\ radii, namely averaged values rather than quantum structures.

In the following, we will perform some pedagogical demonstrations using a density functional inspired by a mass formula.
We write the energy of a nucleus as 
\begin{equation}
E(Z,A) = \int d^3 r \ \rho(r) \ e(\rho(r),\alpha(r)) + \int d^3 r \ f_0 \ |\nabla \rho(r)|^2 + E_{C} \; ,
\label{ldm}
\end{equation}
where the second term represents a  phenomenological description of surface effects, and the Coulomb contribution, $E_C$, is given by    
\begin{equation}
E_{C} = \frac{e^2}{\epsilon_0} \int^{\infty}_{0} d r^{'} \Big (r^{'} \rho_{p}(r^{'}) \int^{r^{'}}_{0} d^3 r \ \rho_{p}(r) \Big ) \; .
\end{equation}  
The parameter $f_0$ is a fitted constant for which we use a value of 65 MeV~fm$^{5}$, consistent with the range determined in Ref.~\cite{Oya2010}. 

We use the two-parameter Thomas-Fermi distribution function to describe the nucleon density:                            
\begin{equation}
\rho (r) = \frac{\rho_{a}}{1 + e^{(r - r_b)/c}} \; .
\end{equation}
The ``radius" $r_b$ and the ``diffuseness" $c$ are themselves evaluated through minimization of the energy per nucleon, Eq.~(\ref{ldm}), while $\rho_a$ is a normalization constant. In Eq.~(\ref{ldm}), $e(\rho(r), \alpha(r))$ is 
the energy per particle in isospin asymmetric matter at some density  $\rho=\rho_n + \rho_p$ (in terms of neutron and proton densities) and isospin asymmmetry $\alpha =\frac{\rho_n-\rho_p}{\rho}$. The main contribution in Eq.~(\ref{ldm}) is of course the energy per particle in asymmetric matter, for which we use the expansion quadratic in $\alpha$:
\begin{equation}
e(\rho, \alpha) \approx e_0 (\rho) + e_{sym}(\rho) \ \alpha^2 \; , 
\label{parab}
\end{equation}
where $ e_0 (\rho) = e(\rho, \alpha=0)$, that is, the SNM EoS ($E/A$ in our figures).
As the EoS is a direct input of the functional, we hope to obtain some insight into the degree to which the energy per nucleon in a nucleus is sensitive to the energy per particle in specific density regions.

\begin{figure*}[!t] 
\centering
\hspace*{-3cm}
\includegraphics[width=8.7cm]{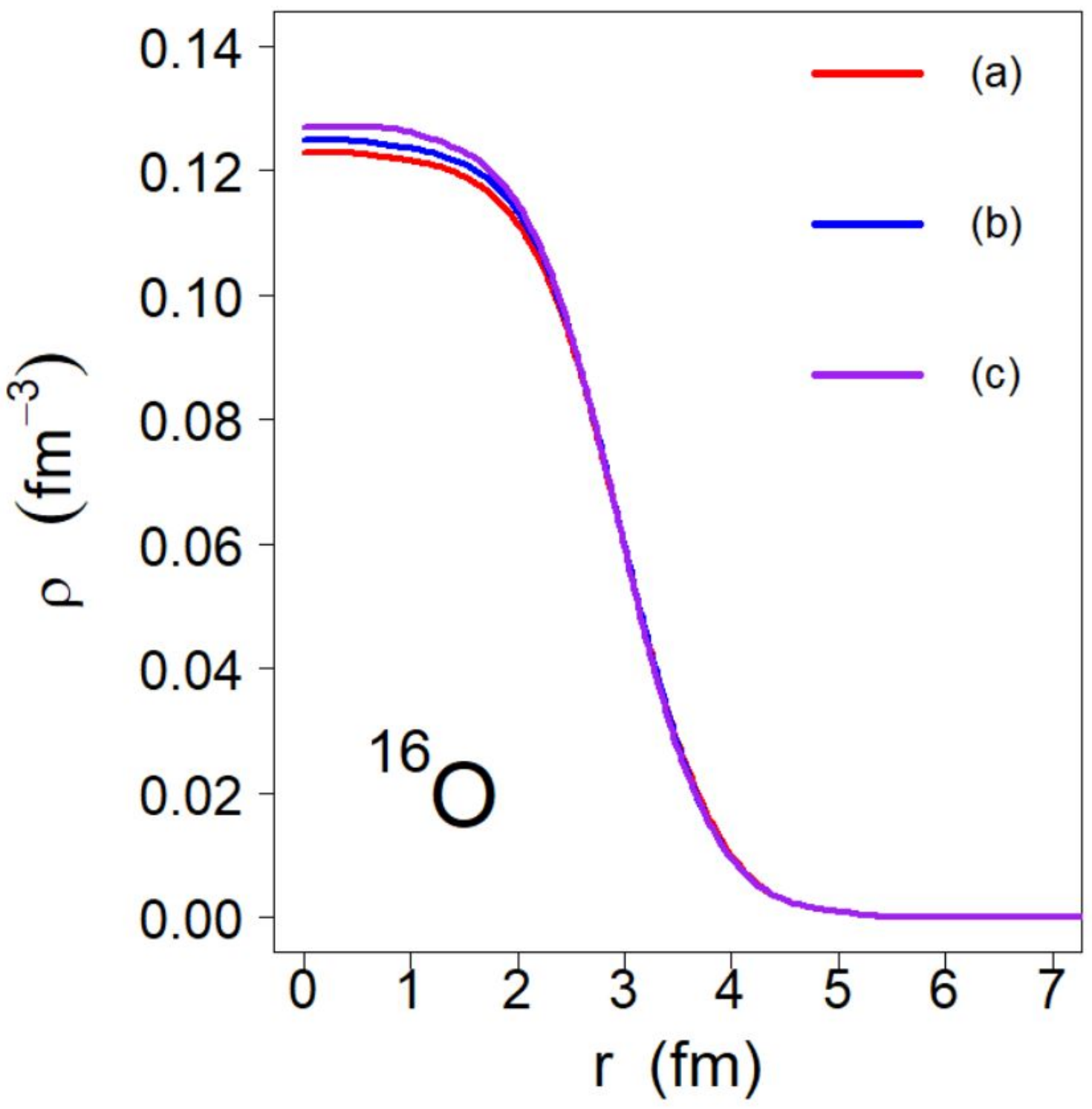}\hspace{0.01in} 
\includegraphics[width=8.7cm]{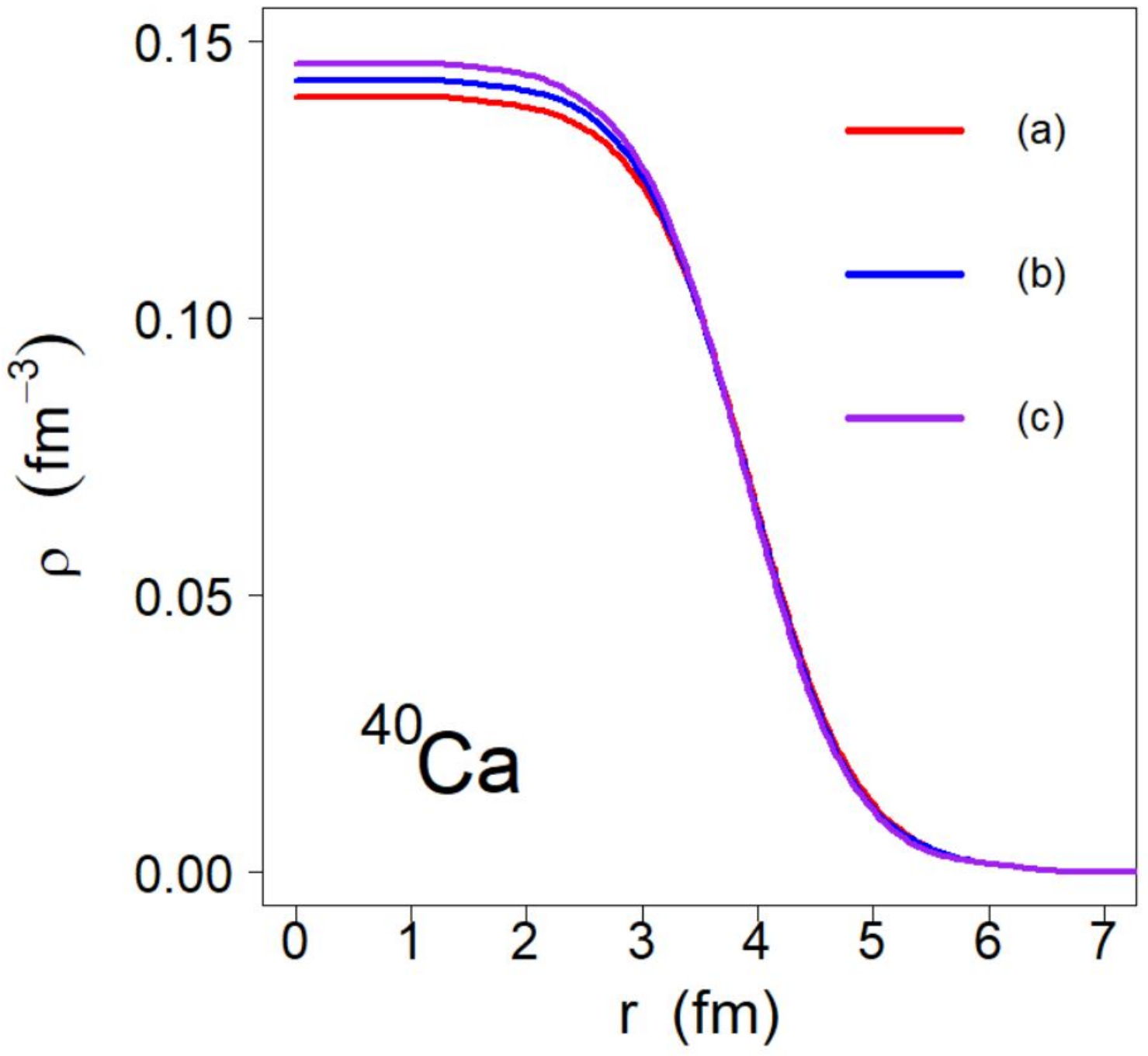}\hspace{0.01in} 
\vspace*{-2.5cm}
 \caption{(Color online) Total matter density distributions as a function of the radial coordinate. Left side: $\rho(r)$ for $^{16}$O obtained as explained in the text, at N$^2$LO, with $\Lambda$=450 MeV and the three $(c_D,c_E)$ combinations indicated as (a), (b), and (c) in Table~\ref{tab1}. Right: same as left, but for $^{40}$Ca.
} 
 \label{rho}
\end{figure*}

Table~\ref{tab1} shows the energy per nucleon and the charge radius for $^{16}$O and $^{40}$Ca which we obtain using in Eq.~(\ref{ldm}) the various EoS shown in Fig.~\ref{dr_all}, along with experimental values. First, we note that our simple intuitive method provides reasonable results, which are consistent with those from Ref.~\cite{Hoppe19}, namely the nuclei are underbound. In Fig.~\ref{rho}, we show the Thomas-Fermi distributions we obtain for some of the cases of Table~\ref{tab1}, which indicates considerably lower central densitiy in $^{16}$O as compared to $^{40}$Ca.

\begin{table*}
\caption{Energy per nucleon and charge radii for selected nuclei. The last two columns show experimental values for the energy per nucleon and the charge radius. Calculations are performed at N$^2$LO. See text for details.}
\label{tab1}
\centering
\begin{tabular*}{\textwidth}{@{\extracolsep{\fill}}ccccccc}
\hline
\hline
Nucleus & $\Lambda$ (MeV) & ($c_{D}$,$c_{E}$)  & E/A (MeV)  & $r_{ch}$ (fm) & E/A (exp.) & $r_{ch}$ (exp.) \\
\hline
\hline
$^{16}$O  & 450 & (a) & -6.8315 & 2.8897 & -7.98    &  2.73  \\ 
          &     & (b) & -6.9196 &  2.8767            &          &         \\ 
          &     & (c) & -7.0125 &  2.8635            &          &            \\ 
          &     &     &         &        &       \\ 
$^{40}$Ca & 450 & (a) & -7.6019 &  3.5736     &  -8.55      &           3.49      \\ 
          &     & (b) & -7.7277 &  3.5537    &              &           \\ 
          &     & (c) & -7.8610 &  3.5330         &                  &              \\ 
          &     &     &         &        &       \\
$^{48}$Ca & 450 & (a) & -7.7565 &  3.6343       &      -8.67     &       3.48    \\
          &     & (b) & -7.8844 &  3.6142        &                &            \\
          &     & (c) & -8.0197 &  3.5930         &               &           \\
		  &     &     &         &        &       \\
$^{208}$Pb& 450 & (a) & -6.8680 &  5.5842           &    -7.87            &     5.50                 \\
          &     & (b) & -7.0254 &  5.5430                        &               &                      \\
          &     & (c) & -7.1930 &  5.5005                          &               &                 \\
\hline
\hline
\end{tabular*}
\end{table*}

As apparent from the EoS discussion, the value of $c_D$ obtained with constraints from $^{16}$O~\cite{Huether+2020} has the effect of lowering the EoS at all densities, as seen from the green curves in Fig.~\ref{n2lo_n3lo} (as well as curve (7) of Fig.~\ref{3nf}), obviously preventing a proper saturating behavior. This suggests that the ideal combination of the $c_D,c_E$-dependent 3NFs should allow for sufficient attraction at densities below saturation, while still enabling saturation.

\begin{figure}[!t] 
\centering
\hspace*{-3cm}
\includegraphics[width=8.7cm]{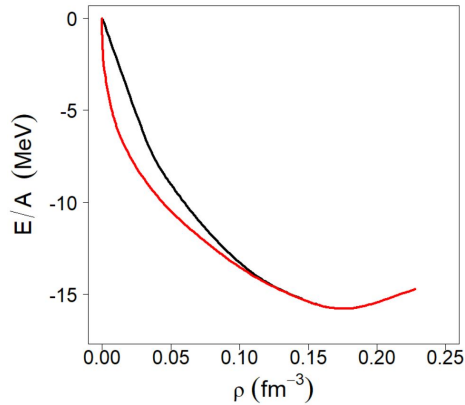}\hspace{0.01in} 
\vspace*{-3.0cm}
 \caption{(Color online) Black curve: EoS ``450 (c)" of Fig.~\ref{dr_all}. Red: {\it ad hoc} modifications below saturation density as described in the text.
}
 \label{adhoc}
\end{figure}

As a demonstration, we take (from Fig.~\ref{dr_all}) the EoS which has a saturating behavior closest to the empirical one, (namely ``450(c)"). We then apply {\it ad hoc} modifications in the density region just below saturation, leaving the regions at or above saturation essentially untouched, until the energy/particle in $^{16}$O shows sufficient binding. The corresponding EoS and results for $^{16}$O and $^{40}$Ca are shown in Fig.~\ref{adhoc} and Table~\ref{adhoc_tab}, respectively. 

\begin{figure}[!t] 
\centering
\hspace*{-3cm}
\includegraphics[width=8.7cm]{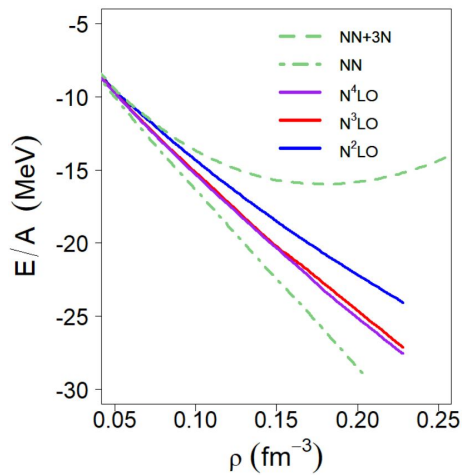}\hspace{0.01in} 
\vspace*{-2.5cm}
 \caption{(Color online) The EoS from Ref.~\cite{magic} for the ``1.8/2.0 (EM)" interaction for 2NF only (dash-dot green) and for 2NF+3NF (dashed green). Furthermore, we show the EoS obtained from the softest (unpublished) 2NFs from the authors of Ref.~\cite{EMN17} at third to fifth order and with  $\Lambda$ = 400 MeV.}
 \label{soft}
\end{figure}

\begin{table*}
\caption{Energy per nucleon and charge radii for $^{16}$O and $^{40}$Ca obtained with modifications of the EoS as demonstrated in Fig.~\ref{adhoc} by the red curve.}
\label{adhoc_tab}
\centering
\begin{tabular*}{\textwidth}{@{\extracolsep{\fill}}ccccc}
\hline
\hline
Nucleus  & E/A (MeV) & $r_{ch}$ (fm) & E/A (exp.) & $r_{ch}$ (exp.) \\
\hline
\hline
$^{16}$O  & -7.99 &  3.03 & -7.98    &  2.73  \\ 
$^{40}$Ca & -8.51 & 3.64    &  -8.55      &           3.49      \\ 
\hline
\hline
\end{tabular*}
\end{table*} 

The $^{16}$O constraint requires more attraction at the lower densities, which is probably why $c_D,c_E$ values corresponding to the green curves in Fig.~\ref{n2lo_n3lo} give good results for binding energies and radii. However, those same values do dramatically lower the EoS of nuclear matter at higher densities. 

We note that, although the binding energy increases, radii become larger. This can be understood as follows: as the binding energy per nuclon increases at the lower (rather than central) densities, the energy minimization (or binding energy maximization) process which we use allows more nucleons to reside outwards, thus enhancing the radius. One must keep in mind that the red curve in Fig.~\ref{adhoc} is just one {\it ad hoc} EoS. An EoS obtained with realistic forces and a $c_D, c_E$ combination consistent with three-nucleon system and nuclear matter constraints might alleviate this problem.

The question is then whether one can ``simulate" the red curve in Fig.~\ref{adhoc} with combinations of realistic 2NFs and 3NFs. One way might be to employ a very soft 2NF and combine it with a repulsive 3NF.

In this context, it is interesting to consider the low-momentum interactions presented in Ref.~\cite{magic}, based on chiral 2NF and 3NF.  Starting from the N$^3$LO 2NF of Ref.~\cite{EM2003}, the authors perform SRG evolution to obtain a (very soft) low-momentum 2NF which is then supplemented by a 3NF fit to the triton binding energy and the point proton radius of $^4$He. We focus here on the interaction that is denoted by ``1.8/2.0 (EM)" in Ref.~\cite{magic}.
 Predictions for finite nuclei are found in Refs.~\cite{Simonis+17,Mor18} and show that the ground-state energies of closed-shell nuclei are well reproduced all the way up to nickel and even the tin isotopes. However, SRG-evolution implies ``induced" 3NFs, which are not included in the interactions of Ref.~\cite{magic}. Thus, this case has some inconsistencies, but might point to the direction of what the 2NF should look like.

So the issue seems to be whether there are (soft) realistic, bare (not SRG evolved) 2NFs that have similar properties as the 2NF of the ``1.8/2.0 (EM)"  interaction of Ref.~\cite{magic}. For this purpose, we display in Fig.~\ref{soft} the EoS predicted by the 2NF and 2NF+3NF of the 1.8/2.0 (EM) interaction. In addition, we show in that figure the nuclear matter predictions for the softest 2NFs constructed by the authors of Ref.~\cite{EMN17} at orders N$^2$LO to N$^4$LO with $\Lambda$ = 400 MeV. As can be seen in Fig.~\ref{soft}, none of these 2NFs are able to generate the attraction in the 2NF of  the ``1.8/2.0 (EM)" interaction of Ref.~\cite{magic}.

In Ref.~\cite{Ekst+18}, chiral nuclear interactions with explicit $\Delta$(1232) degrees of freedom have been  constructed at N$^2$LO. The authors conclude that refined $\Delta$-full interactions have the potential to alleviate the nuclear binding energies {\it vs}. nuclear matter saturation problem also at higher chiral orders. 
On the other hand, while predictions for the radii are satisfactory, binding energies are still underestimated, cf.~Table V of Ref.~\cite{Ekst+18}.  Also, from Fig.~7 of Ref.~\cite{Ekst+18}, SNM is underbound.

\section{Conclusions and outlook}                                                                  
\label{Concl} 

We performed BHF calculations of the nuclear matter EoS with 2N+3N chiral interactions for which ab initio predictions of finite nuclei have been reported. We performed the calculations both at N$^2$LO and at N$^3$LO, and cutoffs of 450 and 500 MeV.

Large differences exist among the values of $c_D, c_E$ fitted through different systems/observables.
However, simultaneously optimal $c_D, c_E$ (especially $c_D$) for few-nucleon systems, medium-mass nuclei, and nuclear matter have so far not been found. It has been the purpose of this paper to shed light on this apparently inherent incompatibility.

For this purpose,  
we investigated the impact  on the EoS of the individual contributions of the leading 3NF and confirmed  that the contribution proportional to $c_D$ is the problematic one.

Further, we determine that the N$^3$LO part of the long-range 3NF is attractive, a result which is parameter independent.

In an effort to gain additional insight on the connection between SNM and nuclei in relation to the fitted values of $c_D, c_E$, we proceeded to calculations of finite nuclei using a mass formula. This simple method, ``fed" directly by the EoS, allowed us to explore the sensitivity of nuclear energies to specific density regions, addressing the question of which densities are mostly probed within a given nucleus. 

From these model calculations, we learned that we need a combination of 2NFs plus 3NFs which allow for sufficient attraction at the lower densities ($\rho \approx 0.07-0.12$ fm$^{-3}$) while still enabling saturation. To achieve this, we can think of two possible scenarios:\\
\underline{Scenario 1}: The 2NF is extremely soft and then combined with a repulsive, strongly density-dependent 3NF contribution. This scenario has been realized by the ``1.8/2.0 (EM)" force of Ref.~\cite{magic} and is the reason for the success of this force. However, as discussed, there are inconsistencies associated with this force and, thus, it does not represent a true solution of the problem. What is needed is a \underline{bare} 2NF of extremely soft character that has so far not been constructed, in spite of vigorous attempts. 
The degree of softness should be such that the resulting EoS has a trend comparable to the red curve in Fig.~\ref{adhoc}. The authors of Ref.~\cite{EMN17} have tried to accomplish this by lowering the cutoff to 400 MeV and using non-local cutoffs, but were not successful (cf. Fig.~\ref{soft}). \\ 
\underline{Scenario 2}: A 2NF of semi-soft character is combined with a 3NF which is attractive at lower densities but repulsive around and above saturation. Since the $c_{1,3,4}$ parameters are fixed from $\pi N$ scattering, the only ``free" parameters to construct such a 3NF are $c_D$ and $c_E$. As demonstrated in Fig.~\ref{3nf}, for positive values of these parameters, their contributions are attractive in SNM.
The $c_D$ contribution is particularly unfortunate because, while it is repulsive in the triton~\cite{Heb15a} and slightly repulsive in SNM at low density, it is very attractive at the higher densities. Therefore, if this contribution is employed to reduce overbinding in the triton, it leads to a disaster in SNM, as demonstrated in Fig.~\ref{3nf}. 
However, as it turns out, for negative values of $c_D$, this contribution is also attractive in SNM and increases strongly with density. This implies that the $c_D$ contribution brings in large off-shell effects. In any case, because of the peculiar nature of the $c_D$-dependent 3NF, it is recommended to keep its parameter small. On the other hand, the $c_E$ parameter acts more regularly, with positive values leading to attraction and negative values causing repulsion, both in the triton and in SNM of any density. There are no off-shell distortions associated with the $c_E$ contribution.

\section*{Acknowledgments}
This work was supported by 
the U.S. Department of Energy, Office of Science, Office of Basic Energy Sciences, under Award Number DE-FG02-03ER41270.

\end{document}